  \documentstyle[12pt]{article}
  
\catcode`@=11
\def\secteqno{\@addtoreset{equation}{section}%
\def\theequation{\thesection.\arabic{equation}}}
\catcode`@=12
\secteqno
\newcommand{\be}{\begin{equation}}
\newcommand{\ee}{\end{equation}}
\newcommand{\bea}{\begin{eqnarray}}
\newcommand{\eea}{\end{eqnarray}}
\newcommand{\bref}[1]{(\ref{#1})}
\newcommand{\nn}{\nonumber}

\newcommand{\slx}{/ {\hskip-0.27cm{x}}}
\newcommand{\slxhs}{/ {\hskip-0.15cm{\hat{x}}}}

\newcommand{\slbL}{/ {\hskip-0.27cm{\bf L}}}

\begin{document}
\thispagestyle{empty}
\vfill
\hfill May 10, 2002\par
\hfill KEK-TH-822\null\par
\vskip 20mm
\begin{center}
{\Large\bf Wess-Zumino term for AdS superstring}\par
\vskip 6mm
\medskip

\vskip 10mm
{\large Machiko\ Hatsuda~~and~~Makoto\ Sakaguchi }\par
\medskip
{\it 
Theory Division,\ High Energy Accelerator Research Organization (KEK),\\
\ Tsukuba,\ Ibaraki,\ 305-0801, Japan} \\
\medskip
E-mail:\ mhatsuda@post.kek.jp,\ 
makoto.sakaguchi@kek.jp
\medskip
\end{center}
\vskip 10mm
\begin{abstract}

\end{abstract} 
We examine a bilinear form Wess-Zumino term
for a superstring in 
 anti-de Sitter (AdS) spaces. 
This is composed of two parts;
a bilinear term in superinvariant currents and 
a total derivative bilinear term 
which is required for the pseudo-superinvariance
of the Wess-Zumino term.
The covariant supercharge commutator containing a string charge is also
obtained. 
\par
\noindent{\it PACS:} 11.30.Pb;11.17.+y;11.25.-w \par\noindent
{\it Keywords:}  Wess-Zumino term; Superalgebra;  Anti-de Sitter
\par\par
\newpage
\setcounter{page}{1}
\parskip=7pt
\section{ Introduction}\par
\indent
Superstring actions in anti-de Sitter spaces 
have been studied in the Green-Schwarz formalism  
\cite{MeTsy,Ram,Berk,HKS,RWS} 
motivated by the AdS/CFT correspondence \cite{Mal}.   
This type of superstring action contains 
a Wess-Zumino term \cite{GSSUST} which is required by the $\kappa$-symmetry  to match 
the number of dynamical degrees of freedom for bosons and fermions \cite{WSkappa}.
The conventional description of the Wess-Zumino term is used in \cite{MeTsy,Ram}
while an alternative description of the Wess-Zumino term, written in a 
bilinear form of superinvariant currents, 
  has been proposed 
for the AdS superstring theories \cite{Berk,RWS}
and for an AdS superstring toy model \cite{HKS}.
The Wess-Zumino term for a superstring in the AdS$_5\times$S$^5$ space
is given in \cite{MeTsy} 
\bea
S_{WZ,{\rm conventional}}=\int d^{3}\sigma ~  H_{[3]}=\int d^2\sigma~B~~,~~
~H_{[3]}=-i\bar{L}\tau_3\slbL L
~~,\label{WZ11}
\eea
while the bilinear form Wess-Zumino term can be given as 
\cite{RWS,KK,HSIW}
\bea
S_{WZ,[2]}=\int d^2\sigma~B_{[2]}~~,~~B_{[2]}=\bar{L}\tau_1 L\label{blWZ}~~
\eea
where $L$ and ${\bf L}$ are superinvariant spinor and vector currents respectively
\footnote{
The charge conjugation matrices are  $C$ and $C'$ which are both anti-symmetric
in the notation \cite{MeTsy}.
We denote 
$\bar{L}_{\alpha\alpha'I}=L^{\beta\beta' I}C_{\beta\alpha}C'_{\beta'\alpha'}$
and~  $\slbL={\bf L}^a\gamma_a+i{\bf L}^{a'}\gamma_{a'}$.}.
Carrying out the integral of the Wess-Zumino term in \bref{WZ11} 
is complicated for AdS cases, so equations of motion and symmetry generators
are hardly obtained except in the light-cone gauge \cite{MeTsylc}.
On the other hand the bilinear form Wess-Zumino term as \bref{blWZ}
is practical for such computations as demonstrated in \cite{HKAdS}.

The difference between $B$ of \bref{WZ11} and $B_{[2]}$ of \bref{blWZ} 
is the supersymmetry property.
$B$ is not superinvariant but pseudo-superinvariant,
while $B_{[2]}$ is manifestly superinvariant.
The pseudo-superinvariance is necessary to give a topological charge 
in the superalgebra as explained in \cite{WitOl,azcTow},
and topological charges classify Wess-Zumino terms \cite{azcTow2}.
A superstring action should contain
a pseudo-superinvariant Wess-Zumino term 
producing a correct topological charge, a string charge.

Now we propose a pseudo-superinvariant two form as a Wess-Zumino term for an AdS superstring
\bea
\tilde{B}=B_{[2]}-d\bar{\theta}~\tau_1~d\theta\label{WZnew},
\eea
which will coincide with the conventional two form $B$ in \bref{WZ11}.
The second term in \bref{WZnew} is a leading term of 
 $B_{[2]}$ under the flat limit,
and it is never obtained from integral of the three form 
$\displaystyle\int_0^1 dt (H_{[3]}\displaystyle|_{\theta\to t\theta})$ 
in  \bref{WZ11}. 
 So this term should be subtracted in order to be $\tilde{B}=B$.
Especially for the computation of topological terms
this subtracted term plays an essential role. 
For computations of local quantities such as equations of motion and
local symmetries,
 this term does not contribute because it is a surface term of an action.
In section 2 in order to show $B=\tilde{B}$
we examine the following criteria \cite{MeTsy}; \par
(a) producing the correct three form gauge field strength, $H_{[3]}$\par 
(b) containing the local $\kappa$ invariance\par
(c) reducing the correct  ``flat limit''~for the IIB superstring action\par\noindent
instead of direct computation by performing integral
of \bref{WZ11}.
The conventional description $B$ gives its flat limit straightforwardly
 \cite{MeTsy},
while naive flat limit makes $B_{[2]}$ to be a trivial total derivative term.
In another word  $B$ in the flat limit is differentiated to be 
 an element of the non-trivial class of 
the Chevalley-Eilenberg cohomology for the supertranslation group, 
while $B_{[2]}$ in the flat limit is differentiated to be zero.
We will show the correct flat limit of the Wess-Zumino term $\tilde{B}$
and suggest the corresponding group contraction which maintains 
the nondegeneracy of the group metric as discussed in \cite{sfet}.

By using the concrete expression of $\tilde{B}$ in \bref{WZnew},
we obtain commutators of the super-AdS$_5\times$S$^5$ symmetry charges
 containing a topological string charge in section 3.
Topological charges in curved backgrounds for 
D-branes
appearing in the supersymmetry commutator
have been calculated by using BPS equations
\cite{CGMP}
and by using Noether method
in the static gauge in the lower order of $\theta$ expansion 
\cite{sato,ohta}. 
The static gauge is not applicable for a
fundamental superstring,
since its zero mode is a superparticle moving along a null geodesics 
in the AdS space \cite{HKAdS} as well as in the flat space. 
The super-AdS$_5\times$S$^5$ algebra
in the light-cone gauge has been examined in
 \cite{MeTsylc} using $B$ 
where a string charge does not show up 
in the light-cone formalism.
In this paper we construct global symmetry charges
by Neother method in canonical language 
with neither any approximation nor any gauge fixing.
A topological string charge in the AdS background is also obtained.
\medskip
\section{ AdS superstring action}\par
\indent

We begin with an action for the AdS superstring
with the Wess-Zumino term \bref{WZnew} given by
\bea
S&=&\displaystyle\int d^2\sigma~{\cal L}
=\displaystyle\int d^2\sigma\left({\cal L}_0+{\cal L}_{WZ}\right)\nn\\
{\cal L}_0&=&-~T
\sqrt{-g}g^{ij}({\bf L}_i^a{\bf L}_{a,j}+{\bf L}_i^{a'}{\bf L}_{a',j})\nn\\
{\cal L}_{WZ}&=&\pm~T
\epsilon^{ij}\left(
\bar{L}_{i}^I(\tau_1)_{IJ} L_j^J
-\partial_i\bar{\theta}^I(\tau_1)_{IJ}\partial_j\theta^J
\right)~~~.
\label{action}
\eea
The notation is the same as the one used by Metsaev and Tseytlin in \cite{MeTsy} \footnote{For AdS$^5\times$S$_5$ case $(a=0,1,...,4,~a'=5,...,9)$ and 
$(\alpha=1,...,4,~\alpha'=1,...,4,~I=1,2)$, 
the super-AdS algebra is given by
\bea
\begin{array}{rcl}
\left[P_a,P_b\right]=J_{ab}&,&\left[P_{a'},P_{b'}\right]=-J_{a'b'}\\
\left[P_a,J_{bc}\right]=\eta_{ab}P_c-\eta_{ac}P_b&,&
\left[P_{a'},J_{b'c'}\right]=\eta_{a'b'}P_{c'}-\eta_{a'c'}P_{b'}\\
\left[J_{ab},J_{cd}\right]=\eta_{bc}J_{ad}+ 3{~\rm terms}&,&
\left[J_{a'b'},J_{c'd'}\right]=\eta_{b'c'}J_{a'd'}+ 3{~\rm terms}\\
\left[Q_I,P_a\right]=\frac{i}{2}Q_{J}\gamma_{a}\epsilon_{JI}&,&
\left[Q_{I},P_{a'}\right]=-\frac{1}{2}Q_J\gamma_{a'}\epsilon_{JI}\\
\left[Q_I,J_{ab}\right]=-\frac{1}{2}Q_{I}\gamma_{ab}&,&
\left[Q_{I},J_{a'b'}\right]=-\frac{1}{2}Q_I\gamma_{a'b'}\\
~~~~~~~~~~~~~~~~~~~~~~~~~~~~~~~~~~~\left\{Q_{\alpha\alpha'I},Q_{\beta\beta'J}\right\}&=&
\delta_{IJ}\left[
-2i{C'}_{\alpha'\beta'}(C\gamma^a)_{\alpha\beta}P_a
+2C_{\alpha\beta}(C'\gamma^{a'})_{\alpha'\beta'}P_{a'}
\right]\label{QQ}\\
&+&
\epsilon_{IJ}\left[
{C'}_{\alpha'\beta'}(C\gamma^{ab})_{\alpha\beta}J_{ab}
-C_{\alpha\beta}(C'\gamma^{a'b'})_{\alpha'\beta'}J_{a'b'}
\right]~~.
\end{array}\nn
\eea
}.
The left-invariant Cartan one-forms of a coset 
\bea
{\rm SU}(2,2|4)/[{\rm SO}(4,1)\times{\rm SO}(5)]~\ni~
G=G(x,\theta)=e^{x P}e^{\theta Q} \nn
\eea
are defined by
\bea
G^{-1}dG&=&{\bf L}^aP_a+{\bf L}^{a'}P_{a'}+\frac{1}{2}{\bf L}^{ab}J_{ab}+\frac{1}{2}{\bf L}^{a'b'}J_{a'b'}
+L^{\alpha\alpha'I}Q_{\alpha\alpha' I}\nn\\
&=&dz^M~L_M^{~~A}~T_A=d\sigma^i~L_i^{~A}~T_A
\label{C1form}
\eea
with $T_A=\{P_a,P_{a'},J_{ab},J_{a'b'},Q_{\alpha \alpha' I}\}$,
$z^M=\{x^m,\theta^\mu\}$ and $\sigma^i=\{\sigma^0,\sigma^1=\sigma\}$
, and 
they are given by
\bea
\left\{\begin{array}{lcl}
{\bf L}^a=e^a+i\bar{\theta}\gamma^a~
    \left(\displaystyle\frac{\sin (\frac{\Psi}{2})}{\Psi/2}\right)^2~D\theta~~~~~~&,&
{\bf L}^{a'}=e^{a'}-\bar{\theta}\gamma^{a'}~
    \left(\displaystyle\frac{\sin (\frac{\Psi}{2})}{\Psi/2}\right)^2~D\theta\\
{\bf L}^{ab}=\omega^{ab}-\bar{\theta}\gamma^{ab}\epsilon~
    \left(\displaystyle\frac{\sin (\frac{\Psi}{2})}{\Psi/2}\right)^2~D\theta&,&
{\bf L}^{a'b'}=\omega^{a'b'}+\bar{\theta}\gamma^{a'b'}\epsilon~
    \left(\displaystyle\frac{\sin (\frac{\Psi}{2})}{\Psi/2}\right)^2~D\theta\\
    L^{\alpha}=\displaystyle\frac{\sin \Psi}{\Psi}~D\theta~~,~~~~~~~~~~~~~~~~~~~~~
   &&\\
   e^a=dx^a+\left(\displaystyle\frac{\sinh x}{x}-1\right)dx^b{\bf \Upsilon}_b^{~~a}
%
&,&
e^{a'}=dx^{a'}+\left(\displaystyle\frac{\sin x'}{x'}-1\right)dx^{b'}{\bf \Upsilon}_{b'}^{~~a'}
%
\\
\omega^{ab}=\frac{1}{2}\left(\displaystyle\frac{\sinh (\frac{x}{2})}{x/2}\right)^2 dx^{[a}x^{b]}
~~~~~~~~~~~~~~~&,&
\omega^{a'b'}=-\displaystyle\frac{1}{2}\left(\displaystyle\frac{\sin (\frac{x'}{2})}{x'/2}\right)^2 dx^{[a'}x^{b']}
\end{array}\right.
\label{LI1}
\eea
where $[ab]=ab-ba$ and
\bea
 D\theta&=&\left[d-\frac{i}{2}\epsilon(\gamma^ae_a+i\gamma^{a'}e_{a'})
    	+\frac{1}{4}(\gamma^{ab}\omega_{ab}+\gamma^{a'b'}\omega_{a'b'})\right]\theta\nn\\ 
 (\Psi^2)^{\alpha\alpha'I}_{~~{\beta\beta' J}}&=&(\epsilon \gamma^{{a}}\theta)^{\alpha\alpha'I}~
(\bar{\theta}\gamma_{{a}})_{\beta\beta' J}-(\epsilon\gamma^{{a'}}\theta)^{\alpha\alpha'I}~
(\bar{\theta}\gamma_{{a'}})_{\beta\beta' J}\nn\\
&&-\frac{1}{2}(\gamma^{{a}{b}}\theta)^{\alpha\alpha'I}~
(\bar{\theta}\gamma_{{a}{b}}\epsilon)_{\beta\beta' J}
+\frac{1}{2}(\gamma^{{a'}{b'}}\theta)^{\alpha\alpha'I}~
(\bar{\theta}\gamma_{{a'}{b'}}\epsilon)_{\beta\beta' J}\nn\\
x&=&\sqrt{x^2}=\sqrt{x^ax_a},~~
x'~=~\sqrt{{x'}^2}=\sqrt{{x}^{a'}x_{a'}}\nn\\
{\bf \Upsilon}_{a}^b&=&\delta_a^{~b}-\displaystyle\frac{x_ax^b}{x^2}~~,
{\bf \Upsilon}_{a'}^{b'}~=~\delta_{a'}^{~b'}-\displaystyle\frac{x_{a'}x^{b'}}{{x'}^2}~~\label{projections}~~~.
\eea

\medskip

\subsection{Three form $H_{[3]}$}

The Cartan one-forms  satisfy the following Maurer-Cartan 
(MC)
 equations
\bea
\left\{\begin{array}{lcl}
dL^{I}&=&\epsilon^{IJ}(\frac{i}{2}{\bf L}^a\gamma_aL^J-\frac{1}{2}{\bf L}^{a'}\gamma_{a'}L^J)
-\frac{1}{4}({\bf L}^{ab}\gamma_{ab}L^I+  {\bf L}^{a'b'}\gamma_{a'b'}L^I)\\
d{\bf L}^a&=&i\bar{L}^I\gamma^a L^I-{\bf L}^b{\bf L}_b^{~a},~~
d{\bf L}^{a'}=-\bar{L}^I\gamma^{a'} L^I-{\bf L}^{b'}{\bf L}_{b'}^{~a'}\\
d{\bf L}^{ab}&=&-{\bf L}^a{\bf L}^b-{\bf L}^{ca}{\bf L}^b_{~c}
-\bar{L}^I\gamma^{ab}\epsilon_{IJ} L^J \\
d{\bf L}^{a'b'}&=&{\bf L}^{a'}{\bf L}^{b'}-{\bf L}^{c'a'}{\bf L}^{b'}_{~c'}
+\bar{L}^I\gamma^{a'b'}\epsilon_{IJ} L^J~~~\end{array}\right.\label{MCeq}~~.
\eea
The first condition (a) is confirmed by 
taking an exterior derivative of $\tilde{B}$ in \bref{WZnew} 
 using the first MC equation and 
symmetric property of indices \footnote{$(C\gamma^{ab})_{\alpha\beta}$ and $(\tau_1)_{IJ}$ are symmetric 
and $C'_{\alpha'\beta'}$ is antisymmetric, while two $L^I$'s are symmetric.
}
\bea
d\tilde{B}&=&d\bar{L}^J(\tau_1)_{JI}L^I+\bar{L}^J(\tau_1)_{JI} dL^L
-d\left(d(\bar{\theta}\tau_1 d\theta)\right)
\nn\\
&=&2 \bar{L}^K(\tau_1)_{KI}
 \{ \epsilon^{IJ}(\frac{i}{2}{\bf L}^a\gamma_aL^J-\frac{1}{2}{\bf L}^{a'}\gamma_{a'}L^J)
-\frac{1}{4}({\bf L}^{ab}\gamma_{ab}L^I+  {\bf L}^{a'b'}\gamma_{a'b'}L^I)\}\nn\\
&=&-i \bar{L}^K(\tau_3)_{KI} \slbL L^I~=~H_{[3]}\label{dBH}~~~~.
\eea 
The result \bref{dBH} is the expected closed three form, $dH_{[3]}=0$.

\medskip

\subsection{{\bf $\kappa$}-invariance}

In order to confirm that the second condition (b),
the $\kappa$-invariance, restricts the coefficient of the Wess-Zumino term,
we set its coefficient to be $b$ as 
\bea
S_{WZ}&=&b\displaystyle\int d^2\sigma~T \tilde{B}
~~\label{blinearB}
\eea
where the surface term does not contribute.
In order to consider arbitrary variations 
$\delta z^M$,
 it is useful to introduce  
\bea
&&~\triangle {\bf L}^AT_A\equiv G^{-1}\delta G=\delta z^M L_M^{~~A}T_A\label{triang}\\
&{\rm e.g.}&~\triangle {\bf L}^a\equiv\delta z^M {\bf L}_M^{~a},
~~\triangle {\bf L}^{ab}\equiv\delta z^M {\bf L}_M^{~ab},
~~\triangle { L}^\alpha\equiv\delta z^M { L}_M^{~\alpha}.\nn
\eea
The important property of the $\kappa$-transformation is given by
\bea
\triangle_\kappa {\bf L}^a&=&
\triangle_\kappa {\bf L}^{a'}~=~
0\nn\\
\triangle_\kappa L^\alpha&=&2~(\slbL ~\kappa)^\alpha~~.
\eea
For a superstring in the general type IIB background the $\kappa$-variations of
Cartan 
one-forms
 are given by 
\bea
\delta_\kappa {L}^\alpha&=&
D({\triangle_\kappa L}^\alpha)
-\frac{1}{4}\triangle_\kappa{\bf L}^{ab}(\gamma_{ab}L)^\alpha
-\frac{1}{4}\triangle_\kappa{\bf L}^{a'b'}(\gamma_{a'b'}L)^\alpha
\label{kappaL}~~\nn\\
\delta_\kappa {\bf L}^a&=&2i\overline{(\triangle_\kappa L)}\gamma^a L
+{\bf L}^b(\triangle_\kappa{\bf L}_b^{~~a})\label{kappaLL}\\
\delta_\kappa {\bf L}^{a'}&=&-2\overline{(\triangle_\kappa L)}\gamma^{a'} L
+{\bf L}^{b'}(\triangle_\kappa{\bf L}_{b'}^{~~a'})\label{kappaLLprime}\nn
\eea
with a covariant derivative $D$
\bea
D(\triangle_\kappa L^{I})&\equiv&d(\triangle_\kappa L^{I})
-\frac{i}{2}\epsilon^{IJ}\slbL(\triangle_\kappa L^{J})
+\frac{1}{4}({\bf L}^{ab}\gamma_{ab}+{\bf L}^{a'b'}\gamma_{a'b'})
(\triangle_\kappa L^\alpha)\nn .
\eea

The $\kappa$ variation of ${\cal L}_0$ is written as
\bea
\delta_\kappa {\cal L}_{0}&=&-T\frac{1}{2}\sqrt{-G}G^{ij}\delta_\kappa G_{ij}
~~,~~G_{ij}={\bf L}_i^a{\bf L}_{a,j}+{\bf L}^{a'}_i{\bf L}_{a',j}
\nn\\
&=&-2Ti\overline{\triangle_\kappa L}~(\sqrt{-G}G^{ij}\slbL_j)~L_i\label{kappaNG}~~.
\eea
On the other hand the $\kappa$-variation of the Wess-Zumino term \bref{blinearB} is given by
\bea
\delta_\kappa {\cal L}_{WZ}&=&2ibT\overline{\triangle_\kappa L}~(\tau_3\epsilon^{ij}\slbL_j)~L_i\label{kappaWZ}~~.
\eea
The factor of the last expression in \bref{kappaNG} is related to the one 
in \bref{kappaWZ} as
\bea
-\sqrt{- G}G^{ij}\slbL_j=\Gamma_{(1)}(\tau_3\epsilon^{ij}\slbL_j)\label{proj}~~
\eea
where
\bea
\Gamma_{(1)}=(\frac{1}{\sqrt{-G}}~\frac{1}{2}\tau_3\slbL)~~
\eea
satisfying 
\bea
~~(\Gamma_{(1)})^2=1~~, {\rm tr}\Gamma_{(1)}=0.
\eea
The $\kappa$-variation of the total action becomes
\bea
\delta_\kappa ({\cal L}_{0}+{\cal L}_{WZ})&=&2Ti\overline{\triangle_\kappa L}~
(\Gamma_{(1)}+b)
(\tau_3\epsilon^{ij}\slbL_j)~L_i~~,
\eea
and the $\kappa$ parameter must satisfy
\bea
&&(\Gamma_{(1)}\pm 1)\kappa =0~~,~~{\rm for} ~b=\pm 1\label{projk}~~,
\eea
where we used the fact $(\Gamma_{(1)}\pm 1)(\triangle_\kappa L)=2\slbL (\Gamma_{(1)}\pm 1)\kappa~$.

\medskip
\subsection{Flat limit}

Now we will examine its ``flat limit'', the third condition (c). 
Under the scaling $x\to (1/R)x$ and $\theta \to (1/\sqrt{R})\theta$, 
the Cartan one-form for $Q$'s \bref{LI1} are expanded in a power series of 
$R$ 
\bea
L^I&=&\displaystyle\sum_{r={\rm half~integer}} \frac{1}{R^r}L^I_{r},\label{L12}
\eea
and especially $L_{1/2}$ and $L_{3/2}$  
\bea
L^I_{1/2}&=&d\theta^^I \label{LRexp}\\
L^I_{3/2}&=&\left[\right.
-\frac{i}{2}(dx^a\gamma_a+idx^{a'}\gamma_{a'})\epsilon^{IJ}\theta^J
+\frac{1}{6}\epsilon^{IJ}(-\gamma^a\theta^J~\bar{\theta}^K\gamma_a d\theta^K
+\gamma^{a'}\theta^J~\bar{\theta}^K\gamma_{a'} d\theta^K)\nn\\
&&+\frac{1}{12}(\gamma^{ab}\theta^I~\bar{\theta}^K\gamma_{ab}\epsilon^{KL}d\theta^L
-\gamma^{a'b'}\theta^I~\bar{\theta}^K\gamma_{a'b'}\epsilon^{KL}d\theta^L)\left.\right]
\eea
are necessary for examining its flat limit.
Corresponding to the expansion of $L$'s,
the two form Wess-Zumino term $B_{[2]}$ is also expanded.  
Its leading term becomes total derivative 
\bea
\frac{1}{{R}}\bar{L}^I_{1/2}(\tau_1)_{IJ}L^J_{1/2}=
\frac{1}{{R}}d\bar{\theta}^I(\tau_1)^{IJ}d{\theta}^J
=\frac{1}{R}d(\bar{\theta}^I{\tau_1}^{IJ}d\theta^J)~~,\label{totald}
\eea 
which is subtracted in our $\tilde{B}$. 
The next to leading term becomes the flat space Wess-Zumino term 
\bea
&&\frac{1}{{R}^2}(\bar{L}^I_{1/2}(\tau_1)_{IJ}L^J_{3/2}+
\bar{L}^I_{3/2}(\tau_1)_{IJ}L^J_{1/2})\nn\\
&&~~~~~
=\frac{1}{R^2}\left[\right.
id\bar{\theta} \tau_3 (dx^a\gamma_a+idx^{a'}\gamma_{a'})\theta\nn\\
&&~~~~~+\frac{1}{3}\{(d\bar{\theta}\gamma^a\tau_3 \theta)(\bar{\theta}\gamma_a d \theta)
-(d\bar{\theta}\gamma^{a'}\tau_3 \theta)(\bar{\theta}\gamma_{a'} d \theta)\}\nn\\
&&~~~~~+\frac{1}{6}\{
(d\bar{\theta}\gamma^{ab}\tau_1 \theta)(\bar{\theta}\gamma_{ab} (i\tau_2) d \theta)
-(d\bar{\theta}\gamma^{a'b'}\tau_1 \theta)(\bar{\theta}\gamma_{a'b'} (i\tau_2)d \theta)
\}\left.\right]~~.\label{Ls1L}
\eea
The cyclic identity of this space is equal to the Jacobi identity of three $Q$'s, 
\bea
{\cal I}_{\alpha\beta\gamma}
&+&{\cal I}_{\beta\gamma\alpha}
~+~{\cal I}_{\gamma\alpha\beta}~=~0\label{cyclic}
\\
{\cal I}_{\alpha\beta\gamma}&=&
{{\cal I}_1}_{\alpha\beta\gamma}
+{{\cal I}_2}_{\alpha\beta\gamma}\nn\\
{\cal I}_{1\alpha\beta\gamma}(\phi)&=&-\delta_{IJ}\{
C'_{\alpha'\beta'}(C\gamma^a)_{\alpha\beta}(\bar{\phi}^L\gamma_a)_{\gamma\gamma'}
-C_{\alpha\beta}(C'\gamma^{a'})_{\alpha'\beta'}(\bar{\phi}^L\gamma_{a'})_{\gamma\gamma'}
\}\epsilon_{LK}\nn\\
{\cal I}_{2\alpha\beta\gamma}(\phi)&=&\frac{1}{2}\epsilon_{IJ}\{
C'_{\alpha'\beta'}(C\gamma^{ab})_{\alpha\beta}(\bar{\phi}^K\gamma_{ab})_{\gamma\gamma'}
-C_{\alpha\beta}(C'\gamma^{a'b'})_{\alpha'\beta'}(\bar{\phi}^K\gamma_{a'b'})_{\gamma\gamma'}
\}\nn
\eea
where $\alpha$ runs $\alpha\alpha'I$ and $\phi$ is an arbitrary spinor.
The second and third terms of the two-form \bref{Ls1L} 
are
expressed in terms of ${\cal I}$ as
\bea
\int~2{\rm nd}+3{\rm rd~{terms}~of}~\bref{Ls1L}&=&-\frac{1}{R^2}\frac{1}{3}\int~
{\cal I}_{\alpha\beta\gamma}(\tau_1 d\theta)~
\theta^\alpha d\theta^\beta \theta^\gamma\nn\\
&=&-\frac{1}{R^2}\frac{1}{3}\int~{\cal I}_{\alpha\beta\gamma}(\tau_1 \theta)~
\{d\theta^\alpha d\theta^\beta \theta^\gamma-\theta^\alpha d\theta^\beta d\theta^\gamma\}
~~,\nn\\
&&\label{B11}
\eea
where a partial integration is performed.
On the other hand the terms are also expressed in terms of ${\cal I}_1$ and ${\cal I}_2$
as
\bea
2{\rm nd}+3{\rm rd~terms
~of}~\bref{Ls1L}&=&\frac{1}{R^2}\frac{1}{3}\{{\cal I}_1(\tau_1  \theta)\theta^\alpha d\theta^\beta d\theta^\gamma
-{\cal I}_2 (\tau_1\theta) \theta^\alpha d\theta^\beta d \theta^\gamma\}
\label{B22} ~~.
\eea
The cyclic identity \bref{cyclic} multiplied with one $\theta$ and two $d\theta$'s
gives following formula
\bea
{\cal I}_{2\alpha\beta\gamma}(2\theta^\alpha d\theta^\beta d\theta^\gamma+d\theta^\alpha d\theta^\beta \theta^\gamma)
=-2{\cal I}_{1\alpha\beta\gamma}\theta^\alpha d\theta^\beta d\theta^\gamma~~.\label{fcy1}
\eea
We will pick up  a
suitable combination of \bref{B11} and \bref{B22}
in such a way that the second and the third terms of 
\bref{Ls1L} 
are
 rewritten to include only ${\cal I}_1$
by using the formula \bref{fcy1}
\bea
&&2{\rm nd}+3{\rm rd~{terms}~of}~\bref{Ls1L}\nn\\
&&~~~~~=
A\bref{B11}+(1-A)\bref{B22}\nn\\
&&~~~~~=\frac{1}{R^2}\{{\cal I}_{1\alpha\beta\gamma} \theta^\alpha d\theta^\beta d \theta^\gamma
-(1-2A){\cal I}_{2\alpha\beta\gamma} \theta^\alpha d \theta^\beta d\theta^\gamma
-A{\cal I}_{2\alpha\beta\gamma} d\theta^\alpha d\theta^\beta\theta^\gamma\}\nn\\
&&~~~~~\stackrel{A=1/4}{=}\frac{1}{R^2}~\frac{1}{3}~\frac{3}{2}{\cal I}_1 \theta^\alpha d\theta^\beta d \theta^\gamma\label{2WZterm}~~.
\eea   
Collecting \bref{2WZterm} and
 other terms in \bref{Ls1L} leads to the following expression of 
 the Wess-Zumino term $\tilde{B}$
\bea
\tilde{B}&=&\bar{L}\tau_1 L-d\bar{\theta}\tau_1 d\theta=
\frac{1}{R^2}(2\bar{L}_{1/2}\tau_1 L_{3/2})
+o(\frac{1}{R^3})
\nn\\
2\bar{L}_{1/2}\tau_1 L_{3/2}&=&
[id\bar{\theta}\tau_3 (dx^a\gamma_a+idx^{a'}\gamma_{a'})\theta\nn\\
&&+\frac{1}{2}\{(\bar{\theta}\tau_3\gamma^a d\theta)(\bar{\theta}\gamma_a d\theta)-
(\bar{\theta}\tau_3\gamma^{a'} d\theta)(\bar{\theta}\gamma_{a'} d\theta)
\}]\nn\\
&=&[
id\bar{\theta}\tau_3 dx^{\hat{a}}\Gamma_{\hat{a}}\theta
+\frac{1}{2}(\bar{\theta}\tau_3\Gamma^{\hat{a}} d\theta)
(\bar{\theta}\Gamma_{\hat{a}} d\theta)]~~~\label{flatB}
\eea 
where $\hat{a}$ runs both $a$ and $a'$.
The $1/R^2$-part 
is the two-form Wess-Zumino term for a superstring in a flat space.
However this $L_{3/2}$ can not be preserved by the conventional IW contraction
where only $L_{1/2}$ is preserved.
In order to define the WZ term consistently even after IW contractions,
new limiting procedure is required.
\par

\medskip
\vskip 6mm
\section{ Super-AdS charges and string charge}\par
\indent

Now we will compute Noether charges for the Super-AdS$_5\times$S$^5$ space
and their commutators.
The total derivative term which is subtracted from the current bilinear term
plays an essential role for the global supersymmetry.
The pseudo invariance of the Wess-Zumino term \bref{WZnew} 
gives a surface term contribution to the supercharge.
Under the global supersymmetry transformation with a parameter $\varepsilon$,
the variation of the Lagrangian \bref{action} comes from only
the subtracted total derivative term
\bea
\delta_\varepsilon {\cal L}&=&\delta_\varepsilon\left(
\mp T\partial_i \epsilon_{ij}
 \bar{\theta}\tau_1\partial_j\theta
\right)\nn\\
&=&\mp T\partial_i \epsilon_{ij}
\left(\delta_\varepsilon \bar{\theta}\tau_1\partial_j\theta
+\bar{\theta}\tau_1\partial_j\delta_\varepsilon \theta\right)\nn\\
&\equiv& \partial_iU^i_\varepsilon\label{UUU}~~~.
\eea
There is no contribution from 
Cartan one forms $L^A$,
since they are invariant under 
the global supersymmetry up to the local Lorentz which 
is cancelled in the Lorentz invariant Lagrangian.
  The supersymmetry charge is written as
\bea
\varepsilon{\cal Q}&=&\displaystyle\int d\sigma
[p\delta_\varepsilon x+\zeta \delta_{\varepsilon}\theta
-U_\varepsilon^0]\label{SUSYcharge}
~~,~~U^0_\varepsilon=\mp 2T\bar{\theta}\tau_1\partial_\sigma(\delta_\varepsilon\theta)
\eea
with $p$ and $\zeta$ being canonical conjugates of $x$ and $\theta$. 
The last term gives the topological string charge in
the superalgebra.

The symmetry transformation rules
$\delta_\varepsilon x$ and $\delta_\varepsilon \theta$ are determined as follows.
Under the supersymmetry transformation an element of a coset ${\cal G}/{\cal H}$ 
is transformed as $G\to   gGh$ with $g\in {\cal G}$ and $h\in {\cal H}$ respectively.
For an infinitesimal global parameter $\varepsilon$, variational one-form 
\bref{triang} is given as  
\bea
G^{-1}\delta_\varepsilon G&=&
G^{-1}~(g-1)~G+(h-1)=\triangle_{\varepsilon} L^AT_A~~\label{llLL}
\\
&=&\delta_{\varepsilon}z^M ~L_M^{~~A}T_A~~.
\nn
\eea 
Therefore 
once $\triangle_\varepsilon L$'s and $L^{-1}$ 
are obtained, symmetry transformation rules are determined as
\bea
\delta_\varepsilon z^M=\triangle_\varepsilon L^A~(L^{-1})_A^{~~M}~~.\label{susyrule}
\eea

At first let us calculate $L^{-1}$ in \bref{susyrule}. 
Coefficients of Cartan one forms are given from \bref{LI1} as
\bea
L_M^{~~A}=\left(
\begin{array}{ccc}
{\bf L}_m^{~~\hat{a}}=e_m^{~~\hat{a}}+\Theta_m^{~~\mu}{\bf L}_\mu^{~~\hat{a}}&
{\bf L}_m^{~~\hat{a}\hat{b}}=\omega_m^{~~\hat{a}\hat{b}}+\Theta_m^{~~\mu}{\bf L}_\mu^{~~\hat{a}\hat{b}}&
{ L}_m^{~~\alpha}=\Theta_m^{~~\mu}{ L}_\mu^{~~\alpha}\\ \\
{\bf L}_\mu^{~~\hat{a}}&{\bf L}_\mu^{~~\hat{a}\hat{b}}&
L_\mu^{~~\alpha}\end{array}
\right)\label{LMAmtrx}
\eea
with
\bea
\left\{\begin{array}{lcl}
e_m^{~~{a}}&=&\delta_m^{~~a}+\left(\displaystyle\frac{\sinh x}{x}-1\right){\bf \Upsilon}_m^{~~a}~,~e_m^{~~{a'}}~=~\delta_m^{~~a'}+\left(\displaystyle\frac{\sin x'}{x'}-1\right){\bf \Upsilon}_m^{~~a'}~\\
\omega_m^{~~{a}{b}}&=&\frac{1}{2}\left(\frac{\sinh (\frac{{x}}{2})}{{x}/2}\right)^2\delta_m^{[{a}}x^{{b}]}~~~~~~~~
,~\omega_m^{~~{a'}{b'}}~=~-\frac{1}{2}\left(\frac{\sin (\frac{{x'}}{2})}{{x'}/2}\right)^2\delta_m^{[{a'}}x^{{b'}]}~\\
\Theta_m^{~\mu}&=&\left\{-\frac{i}{2}\epsilon(e_m^{~{a}}\gamma_{a}+ie_m^{~a'}\gamma_{a'})\theta+\frac{1}{4}(\omega_m^{~~{a}{b}}\gamma_{ab}+\omega_m^{~~a'b'}\gamma_{a'b'})\theta\right\}^\mu\\
{\bf L}_\mu^{~{a}}&=&i\bar{\theta}\gamma^{a}
\left(\frac{\sin (\frac{\Psi}{2})}{\Psi/2}\right)^2~~~~~~~~~~~,~
{\bf L}_\mu^{~{a'}}~=~-\bar{\theta}\gamma^{a'}
\left(\frac{\sin (\frac{\Psi}{2})}{\Psi/2}\right)^2~
\\{\bf L}_\mu^{~{a}{b}}&=&-\bar{\theta}\gamma^{{a}{b}}\epsilon
\left(\frac{\sin (\frac{\Psi}{2})}{\Psi/2}\right)^2~~~~~~~~,~
{\bf L}_\mu^{~{a'}{b'}}~=~\bar{\theta}\gamma^{{a'}{b'}}\epsilon
\left(\frac{\sin (\frac{\Psi}{2})}{\Psi/2}\right)^2\\
L_\mu^{~\alpha}&=&\left(\frac{\sin \Psi}{\Psi}\right)_\mu^{~\alpha}
~~~~~~~~~~~.
\end{array}\right.\label{LAM}
\eea

Its inverse is defined by $L_M^{~~A}(L^{-1})_A^{~~N}=\delta_M^{~~N}$
and is given by 
\bea
(L^{-1})_A^{~~N}=\left(
\begin{array}{cc}
(L^{-1})_{\hat{a}}^{~~n}=(e^{-1})_{\hat{a}}^{~~m}
&(L^{-1})_{\hat{a}}^{~~\nu}\\ \\
(L^{-1})_{\hat{a}\hat{b}}^{~~n}
&(L^{-1})_{\hat{a}\hat{b}}^{~~\nu}\\ \\
(L^{-1})_\alpha^{~~n}&(L^{-1})_\alpha^\nu\end{array}
\right)\label{LAMmtrx}
\eea
with
\bea
\left\{\begin{array}{lcl}
(L^{-1})_a^{~~n}&=&(e^{-1})_{{a}}^{~~n}
~=~\delta_a^{~n}+\left(\displaystyle\frac{x}{\sinh x}-1\right){\bf \Upsilon}_a^{~~n}~,~
(L^{-1})_{ab}^{~~n}=0\\
(L^{-1})_{a'}^{~~n}&=&(e^{-1})_{{a'}}^{~~n}~=~\delta_{a'}^{~n}+\left(\displaystyle\frac{x'}{\sin x'}-1\right){\bf \Upsilon}_{a'}^{~~n}~,~
(L^{-1})_{a'b'}^{~~n}=0\\
(L^{-1})_\alpha^{~n}&=&-i\bar{\theta}\left(
\gamma^{{a}}(e^{-1})_{{a}}^{~~n}+i\gamma^{{a'}}(e^{-1})_{{a'}}^{~~n}
\right)_\beta
\left(\frac{2(\sin (\Psi/2))^2}{\Psi~\sin \Psi}\right)^\beta_{~~\alpha}\\
(L^{-1})_{{a}}^{~~\nu}&=&\frac{i}{2}(\epsilon\gamma_{{a}}\theta)^\nu~~~~~~~~~~,~
(L^{-1})_{{a'}}^{~~\nu}~=~-\frac{1}{2}(\epsilon\gamma_{{a'}}\theta)^\nu~
\\
(L^{-1})_{{a}{b}}^{~~\nu}&=&-\frac{1}{2}(\gamma_{{a}{b}}\theta)^\nu~~~~~~~,~
(L^{-1})_{{a'}{b'}}^{~~\nu}~=~-\frac{1}{2}(\gamma_{{a'}{b'}}\theta)^\nu
\\
(L^{-1})_\alpha^{~~\nu}&=&\left(
\frac{1-2(\sin (\Psi/2))^2}{(\sin \Psi)/\Psi}
\right)_\alpha^{~~\nu}~~~~~~~~~
\end{array}\right.
\label{LMA}~~~
\eea
where this solution is well-defined in a flat limit
although an ambiguity exists caused by the rectangular matirx \bref{LMAmtrx}
and \bref{LAMmtrx} which is removed by 
the local Lorentz degrees of freedom.

Next let us calculate $\triangle_\varepsilon L^A$ for global supersymmetry
 in \bref{susyrule}.~~From the relation of \bref{llLL}, i.e. 
$G^{-1}\delta_\varepsilon G=G^{-1}\varepsilon QG+\frac{1}{2}{\bf h}^{\hat{a}\hat{b}}J_{\hat{a}\hat{b}}$, ~
$\triangle_\varepsilon L^A$'s are obtained as:
\bea
\left\{\begin{array}{lcl}
\triangle_\varepsilon {\bf L} ^{{a}}&=&-2i\bar{\theta}\gamma^{{a}}
\displaystyle\frac{\sin \Psi}{\Psi}\tilde{\varepsilon}~,~~~~~
\triangle_\varepsilon {\bf L} ^{{a'}}~=~2\bar{\theta}\gamma^{{a'}}
\displaystyle\frac{\sin \Psi}{\Psi}\tilde{\varepsilon}
\\
\triangle_\varepsilon {\bf L} ^{{a}{b}}&=&2\bar{\theta}\gamma^{{a}{b}}\epsilon 
\displaystyle\frac{\sin \Psi}{\Psi}\tilde{\varepsilon}+{\bf h}_\varepsilon^{ab}~,~
\triangle_\varepsilon {\bf L} ^{{a'}{b'}}~=~
-2\bar{\theta}\gamma^{{a'}{b'}}\epsilon 
\displaystyle\frac{\sin \Psi}{\Psi}\tilde{\varepsilon}+{\bf h}_\varepsilon^{a'b'}~~
\\
\triangle_\varepsilon L^{\alpha}&=&\cos \Psi \tilde{\varepsilon}~~,~~~~~~~~~~
\tilde{\varepsilon}={\rm exp}[\frac{i}{2}\epsilon ({x}^{{a}}\gamma_a+ix^{a'}\gamma_{a'})]\varepsilon
=e^{i\epsilon \slxhs /2}\varepsilon
 \end{array}\right.\label{llvar}
 \eea
with
\bea
{\bf h}^{ab}&=&-\bar{\theta}\left(\epsilon\gamma^{ab}+
i\gamma^{a}x^b\displaystyle\frac{(\sinh \frac{x}{2}/\frac{x}{2})^2}{\sinh x/x}\right)
\displaystyle\frac{\left(\sin \frac{\Psi}{2}/\frac{\Psi}{2}\right)^2}{\sin \Psi/\Psi}
\tilde{\varepsilon}\nn\\
{\bf h}^{a'b'}&=&\bar{\theta}\left(\epsilon\gamma^{a'b'}-
\gamma^{a'}x^{b'}\displaystyle\frac{(\sin \frac{x'}{2}/\frac{x'}{2})^2}
{\sin x'/x'}\right)
\displaystyle\frac{\left(\sin \frac{\Psi}{2}/\frac{\Psi}{2}\right)^2}{\sin \Psi/\Psi}
\tilde{\varepsilon}\label{bfh}~~.
\eea
  Combining \bref{susyrule}, \bref{LMA}, \bref{llvar} and \bref{bfh} gives 
 supertransformation rules as
 \bea
\left\{\begin{array}{lcl}
 \delta_{\varepsilon} x^m&=&-i\bar{\theta}\gamma^m
 \displaystyle\frac{\left(\sin \frac{\Psi}{2}/\frac{\Psi}{2}\right)^2}{\sin \Psi/\Psi}
 \tilde{\varepsilon}
\\
\delta_\varepsilon \theta^\mu&=&
\displaystyle\frac{\Psi}{\sin \Psi}\tilde{\varepsilon}\\
&+&
\frac{1}{4}\left[\gamma_{ab}\theta
\bar{\theta}\left(\epsilon \gamma^{ab}+i\gamma^c\omega_c^{~ab}
\right)
-\gamma_{a'b'}\theta
\bar{\theta}\left(\epsilon \gamma^{a'b'}+\gamma^{c'}\omega_{c'}^{~a'b'}
\right)
\right]
\displaystyle\frac{\left(\sin \frac{\Psi}{2}/\frac{\Psi}{2}\right)^2}
{\sin \Psi / \Psi}\tilde{\varepsilon}
\end{array}\right. \label{susy}
 \eea 
 with
 \bea
 \gamma^m&=&\left\{\begin{array}{l}
  \gamma^a{(e^{-1})}_{a}^{~~m} ~,~~{\rm for}~m=0\sim 4\\
 i\gamma^{a'}{(e^{-1})}_{a'}^{~~m}~,~~{\rm for}~m=5\sim  9
 \end{array}\right.~~\nn\\
\omega_c^{ab}&=&{(e^{-1})}_{c}^{~~m}\omega_m^{~~ab}=\delta_c^{~[a}x^{b]}\frac{1}{2}
\displaystyle\frac{\left(\sinh \frac{x}{2}/\frac{x}{2}\right)^2}
{\sinh x /x}\nn\\
\omega_{c'}^{a'b'}&=&{(e^{-1})}_{c'}^{~~m}\omega_m^{~~a'b'}
=-\delta_{c'}^{~[a'}x^{b']}\frac{1}{2}
\displaystyle\frac{\left(\sin \frac{x'}{2}/\frac{x'}{2}\right)^2}
{\sin x' /x'}~~~.
 \eea
Inserting \bref{susy} into \bref{SUSYcharge}, 
the supersymmetry charge is written as
\bea
&&{\cal Q}_{{\alpha}\alpha' I}\nn\\
&&
=
\displaystyle\int d\sigma \left[
-i\bar{\theta}(\gamma^ap_a+i\gamma^{a'}p_{a'})
\displaystyle\frac{\left(\sin\frac{\Psi}{2}/\frac{\Psi}{2}\right)^2
}{\sin\Psi/\Psi}e^{i\epsilon \slxhs /2}\right.
\label{Q}
\\
&&+
\zeta\left\{1
+\frac{1}{4}\left(\gamma_{ab}\theta\bar{\theta}
(\epsilon\gamma^{ab}+i\gamma^c\omega_c^{ab})
-\gamma_{a'b'}\theta\bar{\theta}
(\epsilon\gamma^{a'b'}+\gamma^{c'}\omega_{c'}^{a'b'})\right)
\left(\displaystyle\frac{\sin\frac{\Psi}{2}}{\Psi/2}\right)^2
\right\}
\displaystyle\frac{1}{\sin \Psi/\Psi}e^{i\epsilon \slxhs /2}\nn\\
&&\left.\pm 2T\bar{\theta}\tau_1\partial_\sigma
\left\{\left(1+\frac{1}{4}\left(\gamma_{ab}\theta\bar{\theta}
(\epsilon\gamma^{ab}+i\gamma^c\omega_c^{ab})
-\gamma_{a'b'}\theta\bar{\theta}
(\epsilon\gamma^{a'b'}+\gamma^{c'}\omega_{c'}^{a'b'})\right)
\left(\displaystyle\frac{\sin\frac{\Psi}{2}}{\Psi/2}\right)^2
\right)\displaystyle\frac{1}{\sin \Psi/\Psi}e^{i\epsilon \slxhs /2}
\right\}
\right]\nn
\eea

The momentum charge and the Lorentz charge are  
analogously obtained in the appendix and given by
 \bea
{\cal P}_{\hat{a}}&=&
\displaystyle\int d\sigma~\left[
p_{\hat{a}} +p_{\hat{b}}x_{\hat{c}}\omega_{\hat{a}}^{~\hat{b}\hat{c}}
+\zeta\gamma_{\hat{b}\hat{c}}\theta\omega_{\hat{a}}^{~\hat{b}\hat{c}}
\right]~,~~~{\rm for}~\hat{a},\hat{b},\hat{c}=0\sim 4~{\rm or}~
5\sim 9\label{P}\nn
\\ 
 {\cal J}_{\hat{a}\hat{b}}&=&
\displaystyle\int d\sigma~[p_{[\hat{a}}x_{\hat{b}]}+\frac{1}{2}\zeta\gamma_{\hat{a}\hat{b}}\theta]~~~
\label{J}~~~.
\eea

These charges of the super-AdS$_5\times$S$^5$ space 
satisfy the following commutators
\bea
\left\{{\cal Q}_{\alpha\alpha'I},{\cal Q}_{\beta\beta'J}\right\}&=&
-2i\delta_{IJ}(CC'\gamma^{\hat{a}})_{\alpha\alpha'\beta\beta'}{\cal P}_{\hat{a}}
+\epsilon_{IJ}(CC'\gamma^{\hat{a}\hat{b}})_{\alpha\alpha'\beta\beta'}{\cal J}_{\hat{a\hat{b}}}~\label{QQalg}\\
&&-2i(\tau_3)_{IJ}(CC'\gamma^{\hat{a}})_{\alpha\alpha'\beta\beta'}{\cal Z}_{\hat{a}}
~\nn
\eea 
\bea
&&\left[{\cal P}_{a},{\cal P}_b\right]={\cal J}_{ab}~,~\left[{\cal P}_{a'},{\cal P}_{b'}\right]=-{\cal J}_{a'b'}~\nn\\
&&\left[{\cal P}_{\hat{a}},{\cal J}_{{\hat{b}}{\hat{c}}}\right]=\eta_{{\hat{a}}{\hat{b}}}{\cal P}_{\hat{c}}-\eta_{{\hat{a}}{\hat{c}}}{\cal P}_{\hat{b}}
~~,~~\left[{\cal J}_{{\hat{a}}{\hat{b}}},{\cal J}_{{\hat{c}}\hat{d}}\right]=\eta_{{\hat{b}}{\hat{c}}}{\cal J}_{{\hat{a}}\hat{d}}+ 3{~\rm terms}
\label{sAdS}\\
&&\left[{\cal Q}_I,{\cal P}_{a}\right]=\frac{i}{2}{\cal Q}_{J}\gamma_{a}
\epsilon_{JI}~~,~~
\left[{\cal Q}_I,{\cal P}_{a'}\right]=
-\frac{1}{2}{\cal Q}_{J}\gamma_{a'}\epsilon_{JI}
~~,~~\left[{\cal Q}_I,{\cal J}_{{\hat{a}}{\hat{b}}}\right]=-\frac{1}{2}{\cal Q}_{I}\gamma_{{\hat{a}}{\hat{b}}}\nn
\eea
up to the local Lorentz generator. 
The topological term ${\cal Z}$~is obtained from the surface term 
as explained in the beginning of this section
\bea
&&\left[\delta_{\varepsilon'}(-\displaystyle \int d\sigma~U^0_\varepsilon)
-(\varepsilon\leftrightarrow\varepsilon')\right]\displaystyle{|}_{\theta=0}\nn\\
&&=\varepsilon'^{(\hat{\beta}J}\varepsilon^{\hat{\alpha}I)}
(\pm 2T)\left(CC'e^{-i\epsilon\slxhs /2} ~\tau_1~\partial_\sigma(e^{i\epsilon\slxhs /2})
\right)_{\hat{\alpha}I\hat{\beta}J}\nn\\
&&=\varepsilon'^{(\hat{\beta}J}\varepsilon^{\hat{\alpha}I)}
(\pm T)(-\tau_3)\partial_\sigma(iCC'\slx~\displaystyle\frac{\sinh X}{X})\nn\\
&&=\varepsilon'^{(\hat{\beta}J}\varepsilon^{\hat{\alpha}I)}
i(\tau_3)_{IJ}(CC'\gamma^{\hat{a}})_{\hat{\alpha}\hat{\beta}}{\cal Z}_{\hat{a}}
\eea
where $X=\sqrt{x^ax_a+x^{a'}x_{a'}}$ and
\bea
{\cal Z}_{\hat{a}}&=&
\pm T\displaystyle\int d\sigma~\partial_\sigma
\left(x_{\hat{a}}~\displaystyle\frac{\sinh X}{X}\right)
~~.\label{Fcharge}
\eea 
For a zero-mode of the string this string charge vanishes
and 32 supersymmetries remain.
For massive excited states this string charge breaks 
half supersymmetries as same as BPS states.  
After rescaling $x\to x/R$ and taking the flat limit $R\to \infty$
this reduces into the flat string charge,
$T\displaystyle\int d\sigma~\partial_\sigma x_{\hat{a}}$.
 
\medskip

\section{ Conclusions and discussions}\par
\indent
We have shown that the pseudo-superinvariant Wess-Zumino term written in the form of
$\tilde{B}$ in \bref{WZnew}
satisfies the three conditions, (a) correct three form $H_{[3]}$, 
(b) $\kappa$-invariance and (c) correct flat limit. 
We have constructed global charges of the super-AdS space ${\cal Q},{\cal P},{\cal J}$ and 
a string charge ${\cal Z}$ in \bref{Fcharge} 
appeared in the supercharge commutator.
 We have also mentioned that 
a generalization of the IW contraction is required 
 to give a correct flat limit of the bilinear Wess-Zumino term,
 where the next leading term in the limiting procedure 
 is preserved to make the fermionic part of the group metric to be nondegenerate,
 i.e. not only $L_{1/2}$ but also $L_{3/2}$ should be preserved in \bref{L12}
 as shown in \bref{flatB}.
 In a flat space the bilinear Wess-Zumino term can not exist 
as shown in \cite{azcTow}, but 
in AdS spaces the bilinear form Wess-Zumino terms 
exist. This fact is a reflection of the fact that
the super-AdS algebra is nondegenerate.
In order to have the bilinear Wess-Zumino term even in a flat space
after some group contraction, we need  a
generalization of the IW-contraction where the scale parameter
does not disappear completely and the resultant superalgebra is
nondegenerate \cite{HSIW}.

Once AdS brane actions are obtained,
pp brane actions can be easily obtained by an analogous limiting procedure to 
the section 2.3 corresponding to the contraction of super-AdS groups \cite{HKSpp}.

\vskip 6mm
{\bf Acknowledgments}\par
We are grateful to Kiyoshi Kamimura for pointing out the equation \bref{susyrule}
  and for fruitful discussions.
We thank a referee for clarifying the point 
 ``Are $B$ and $B_{[2]}$ same or not?" to improve our previous paper \cite{HSIW}.  

\vskip 10mm
\appendix
\section{Appendix}\par
The momentum and Lorentz charges are also obtained analogously to
the supersymmetry charges by taking 
 infinitesimal parameters
 $y^{\hat{a}}$ and 
 $\lambda^{\hat{a}\hat{b}}=\{\lambda^{ab}, ~\lambda^{a'b'}\}
 $. 
 The momentum charge is given by
 \bea
y^{\hat{a}}{\cal P}_{\hat{a}}=\displaystyle\int d\sigma
[p_m \delta_y x^m+\zeta_\mu \delta_y\theta^\mu]~~.
 \eea
Translation variation leads to $\triangle_y L^A$  in \bref{susyrule}
as
 \bea
\left\{\begin{array}{lcl}
\triangle_y {\bf L}^{{a}}&=&\tilde{y}^a
-i\bar{\theta}\gamma^{{a}}
\displaystyle\frac{\sin \Psi}{\Psi}\tilde{Y}\theta~,~
\triangle_y {\bf L}^{{a'}}~=~\tilde{y}^{a'}+\bar{\theta}\gamma^{{a'}}
\displaystyle\frac{\sin \Psi}{\Psi}\tilde{Y}\theta
\\
\triangle_y {\bf L}^{{a}{b}}&=&\tilde{\tilde{y}}^{ab}+
\bar{\theta}\gamma^{{a}{b}}\epsilon 
\displaystyle\frac{\sin \Psi}{\Psi}\tilde{Y}\theta+{\bf h}_y^{ab}~,~
\triangle_y {\bf L}^{{a'}{b'}}~=~\tilde{\tilde{y}}^{a'b'}-
\bar{\theta}\gamma^{{a'}{b'}}\epsilon 
\displaystyle\frac{\sin \Psi}{\Psi}\tilde{Y}\theta+{\bf h}_y^{a'b'}~,~
\\
\triangle_y L^{\alpha}&=&\displaystyle\frac{\sin \Psi}{\Psi}\tilde{Y}\theta~~\end{array}\right.\label{lly}
\eea
with
\bea
\left\{\begin{array}{lcl}
\tilde{Y}\theta&=&
[-\frac{i}{2}\epsilon(\tilde{y}^a\gamma_a+i\tilde{y}^{a'}\gamma_{a'})
+\frac{1}{4}(\tilde{\tilde{y}}^{ab}\gamma_{ab}+\tilde{\tilde{y}}^{a'b'}\gamma_{a'b'})]\theta
\\
\tilde{y}^a&=&y^a+(\cosh x-1)y^b{\bf \Upsilon}_b^{~~a}~,~
\tilde{y}^{a'}~=~y^{a'}+(\cos x'-1)y^{b'}{\bf \Upsilon}_{b'}^{~~a'}~\\
\tilde{\tilde{y}}^{ab}&=&-x^{[a}y^{b]}\frac{\sinh x}{x}~,~
\tilde{\tilde{y}}^{a'b'}~=~x^{[a'}y^{b']}\frac{\sin x'}{x'}~
 \end{array}\right.\label{llyt}
 \eea
The subgroup parameter ${\bf h}$'s are determined by 
\bea
{\bf h}_{y}^{ab}=\delta_{y}x^m{\bf L}_m^{~ab}-\triangle_y {\bf L}^{ab}~~,~~
{\bf h}_{y}^{a'b'}=\delta_{y}x^m{\bf L}_m^{~a'b'}-\triangle_y {\bf L}^{a'b'}
\eea
where $\delta_y x^m$ is determined independently on ${\bf h}$'s because of
$(L^{-1})_{\hat{a}\hat{b}}^{~~m}=0.$
Using the above relations the transformation rules under the translation are 
obtained analogously to \bref{susyrule} by
\bea
\left\{\begin{array}{lcl}
 \delta_{y} x^m&=&\left\{\begin{array}{l}
 y^m+(\cosh x-1) y^n {\bf \Upsilon}_n^{~m},~~~{\rm for}~m,n=0\sim 4\\
 y^m+(\cos x'-1)y'^n{\bf \Upsilon}_n^{~m},~~~{\rm for}~m,n=5\sim 9\end{array}\right.\\
\delta_y \theta^\mu&=&\left(\frac{1}{4}\gamma_{ab}\theta
\omega_m^{~~ab}+\frac{1}{4}\gamma_{a'b'}\theta
\omega_m^{~~a'b'}\right)y^m
\end{array}\right. \label{susyy}~~.
 \eea

 The Lorentz charge is given by
\bea
\frac{1}{2}\lambda^{\hat{a}\hat{b}}{\cal J}_{\hat{a}\hat{b}}=
\displaystyle\int d\sigma[p_m \delta_\lambda x^m+\zeta_\mu \delta_\lambda\theta^\mu]~~.
\eea
The Lorentz variation leads to
 $\triangle_\varepsilon L^A$  in \bref{susyrule}
 obtained as \bref{lly} where  $\tilde{y}^{\hat{a}}$, $\tilde{\tilde{y}}^{\hat{a}\hat{b}}$,
 $\tilde{Y}\theta$ and the subscript $_y$
are replaced by
  $\tilde{\tilde{\lambda}}^{\hat{a}}$, $\tilde{\lambda}^{\hat{a}\hat{b}}$,
 $\tilde{\Lambda}\theta$ and a subscript $_\lambda$
with
\bea
\left\{\begin{array}{lcl}
\tilde{\Lambda}\theta&=&[-\frac{i}{2}\epsilon(\tilde{\tilde{\lambda}}^a\gamma_a+i\tilde{\tilde{\lambda}}^{a'}\gamma_{a'})+\frac{1}{4}(\tilde{\lambda}^{ab}\gamma_{ab}+\tilde{\lambda}^{a'b'}\gamma_{a'b'})]\theta\\
\tilde{\tilde{\lambda}}^a&=&\lambda^{a}_{~~b}x^b\displaystyle\frac{\sinh x}{x}~,~\tilde{\tilde{\lambda}}^{a'}~=~\lambda^{a'}_{~~b'}x^{b'}\displaystyle\frac{\sin x'}{x'}\\
\tilde{\lambda}^{ab}&=&\lambda^{ab}+x^{[a}\lambda^{b]}_{~c}x^c\displaystyle\frac{\cosh x-1}{x^2}~,~
\tilde{\lambda}^{a'b'}~=~\lambda^{a'b'}+x^{[a'}\lambda^{b']}_{~c}x^c
\displaystyle\frac{\cos x'-1}{{x'}^2}~
 \end{array}\right.\label{lllam}~~~.
 \eea
The Lorentz variation rules are obtained as
\bea
\left\{\begin{array}{lcl}
 \delta_{\lambda} x^m&=& \lambda^m_{~~n}x^n\\
\delta_{\lambda} \theta^\mu&=&\frac{1}{4}(\gamma_{ab}\lambda^{ab}
+\gamma_{a'b'}\lambda^{a'b'})\theta
\end{array}\right. \label{susyl}~~.
 \eea


\end{document}